\documentclass[fleqn,12pt]{wlscirep}
\usepackage{amsmath}
\usepackage{graphicx}
\pdfoutput=1
\usepackage{footmisc}
\usepackage{color}

\begin{document}

\title {Spontaneous breaking of time-reversal symmetry in topological superconductors}

\author[1]{Igor N. Karnaukhov}
\affil[1]{G.V. Kurdyumov Institute for Metal Physics, 36 Vernadsky Boulevard, 03142 Kiev, Ukraine}
\affil[*]{karnaui@yahoo.com}

\begin{abstract}
We study the behavior of spinless fermions in superconducting state, in which the phases of the superconducting order parameter depend on the direction of the link. We find that the energy of the superconductor depends on the phase differences of the superconducting order parameter. The solutions for the phases corresponding to the energy minimuma, lead to a topological superconducting state with the nontrivial Chern numbers. We focus our quantitative analysis on the properties of topological states of superconductors with different crystalline symmetry  and show that the phase transition in the topological superconducting state is result of spontaneous breaking of time-reversal symmetry in the  superconducting state. The peculiarities in the  chiral gapless edge modes behavior are studied, the Chern numbers are calculated.
\end{abstract}

\maketitle

\section*{Introduction}

Within a mean-field treatment of the problem, the superconducting state is characterized by the pairing of a macroscopic number of electrons with opposite spins. The superconducting ground state can be thought of as the condensation of a macroscopic number of such bosons, where they all have the same phase.  The real superconducting  order parameter breaks the U(1) symmetry and maintains the time-reversal symmetry. At the same time there is no absolute value for the phase of the wave function of a single piece of superconductor in free space: any phase can be chosen between 0 and $2\pi$. The phase coherent tunneling across a junction between two superconductors implies a persistent current determined by the phase difference of the superconducting order parameters with the $2 \pi$ periodicity. The Josephson effect has been considered in refs \cite{1,2} in the framework of the Kitaev model \cite{4} (see also \cite{3a}).

The phase of the superconducting order parameter is not fixed in superconducting state, it does not break symmetry of the Hamiltonian (the phase difference is fixed in fact). When phases of superconducting order parameters of interacting superconductors are taken into account the time reversal symmetry is broken. The realization of the topological states of the system depends on the presence or absence of particle-hole, time-reversal symmetries. We will show that a new order parameter, determined by the phase differences of the superconducting order parameter, spontaneously breaks  the time-reversal symmetry and leads to the topological superconducting state. This scenario is suggested as a possible mechanism for realization of topological superconductors (TSCs).

Different mechanisms for realization of TSCs have been proposed \cite{TSC1,TSC2,TSC3,TSC4,Roy, B}. The superconductors with  chiral $(d+id)$, $(p+ip)$ order-parameters exhibit Majorana vortex bound states and gapless chiral edge modes, which carry spin currents \cite{TSC5,TSC6,TSC7,TSC8}. The TSC with the $(p+ip)$-pairing of spinless fermions in two-dimension, which has chiral Majorana fermion states propagating along the edges, has been considered in  \cite{TSC5a}. The standard method for realizing topological states  reduces to considering the symmetry breaking of the system due to the presence of additional interaction \cite{KM,Hof,A,Hal}. The topological state are described in the framework of topological band theories \cite{10,11} (as a rule without interaction), which characterize a class of topological insulator or TSC by a topological invariant.

We demonstrate that due to stable solutions for phases of the superconducting order parameter (which acts as a new order parameter), spontaneous breaking  of  time-reversal symmetry  leads to topological state of superconductors. A similar approach for describing the states of a topological insulator, realized on a hexagonal lattice, was proposed in \cite{K2}. In other words the continuous gauge symmetry is spontaneously broken by the stable thermodynamic state, the state of the system selects a global difference phases of the superconducting order parameter. The absence of any required interaction or external field is, of course, an experimental simplification in the investigation of topological states.

\begin{figure}[tbp]
     \includegraphics[width=\linewidth]{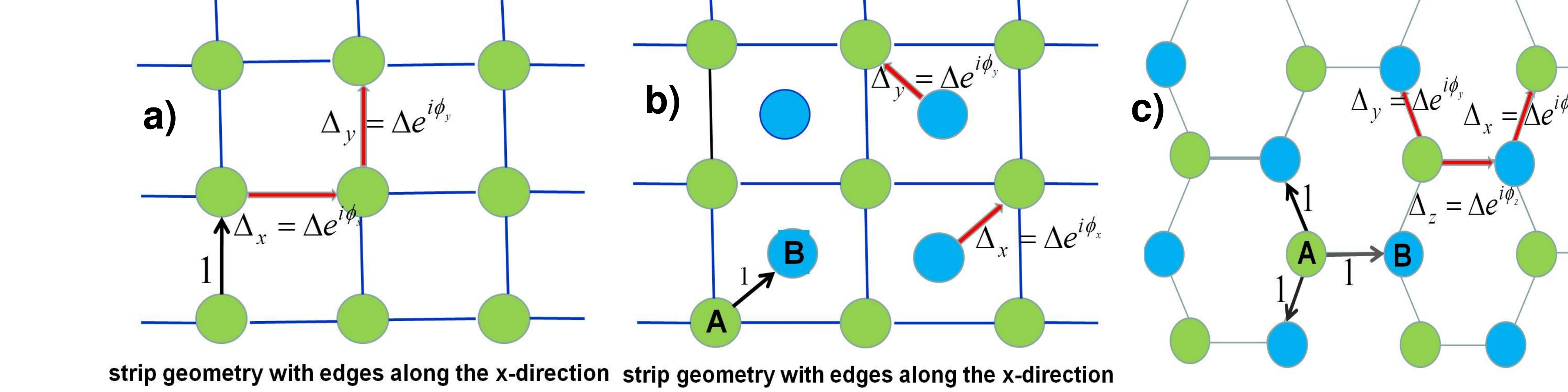}
    \caption{(Color online) Square a), centered square b) and hexagonal c) lattices and the parameters of the model (hopping integrals
and superconducting order parameters).
    The types of links on the lattices:
    $1$ denotes the magnitude of the nearest-neighbor hopping integral, $\Delta_x$, $\Delta_y$ (in square) and
$\Delta_x$, $\Delta_y$, $\Delta_z$ (in hexagonal) lattices determine the pairing of spinless fermions located at
the nearest-neighbor lattice sites along the links. The unit cells of the centered square and hexagonal lattices
contain two atoms A and B. A strip geometry with edges is fixed along the \emph{x}-direction for the square lattices.
 } \label{fig:Model}
\end{figure}

\section*{Model}

We will analyze the behavior of spinless fermions on the 2D lattices in the framework of the model of chiral $p$-wave superconductors \cite{4}, We will only consider the short-range nearest-neighbor superconducting pairing, $p-$ pairing between spinless fermions in the nearest-neighbor pairs of sites. The concerned Hamiltonian $ {\cal H}= {\cal H}_0+ \Delta {\cal H}$ is written on the square and hexagonal (honeycomb) lattices as
\begin{equation}
 {\cal H}_0= - \sum_{<ij>}a^\dagger_{i}a_j - 2\mu \sum_{j} n_j,\\
 \Delta {\cal H}=\Delta_x \sum_{x-links} a^\dagger_{i}a^\dagger_{j}+\Delta_y \sum_{y-links} a^\dagger_{i}a^\dagger_{j}+ \Delta_z \sum_{z-links} a^\dagger_{i}a^\dagger_{j}+h.c. ,
        \label{eq-H1}
\end{equation}
where $a^\dagger_{j} $ and $a_{j}$ are the spinless fermion operators on a site \emph{j} with the usual anticommutation relations,  $n_j$ denotes the density operator. The Hamiltonian (\ref{eq-H1}) describes the hoppings of spinless fermions between the nearest-neighbor lattice sites  with the magnitudes equal to unit (see in Figs \ref{fig:Model}), $\mu$ is the chemical potential. Other terms represent pairing terms with superconducting order parameter $\Delta_{x,y,z}$, which is determined along the link of different types: the $x-,y-$ links in the square lattices and  the $x-,y-,z-$ links in the hexagonal lattice (see in Figs \ref{fig:Model}),
$\Delta_x=\Delta \exp(i\phi_x)$, $\Delta_y=\Delta\exp(i\phi_y)$, $\Delta_z=\Delta\exp(i\phi_z)$,
the components of the superconducting order parameter have different phases $\phi_x$, $\phi_y$, $\phi_z$ and $\Delta>0$ (see in Figs \ref{fig:Model}).
The phase differences $\phi=\phi_x- \phi_y$ in square and $\phi= \phi_x-\phi_z$, $\varphi= \phi_y-\phi_z$  in hexagonal and cubic lattices have a physical meaning. We consider a family of the two parameter models, in which the absolute value of the superconducting order parameter has the same value and does not differ in strength. We show that the nontrivial solutions $\phi= \pm \pi/2$ lead to \emph{p+ip} superconducting state on a square lattice. The model Hamiltonian is convenient to determine in the  real space \cite{3,4}, we ignore spin of the fermions and focus on the relatively simpler case of TSCs.

\section*{TSC on a square lattice}

\begin{figure}[tp]
   \centering{\leavevmode}
 \includegraphics[width=.65\linewidth]{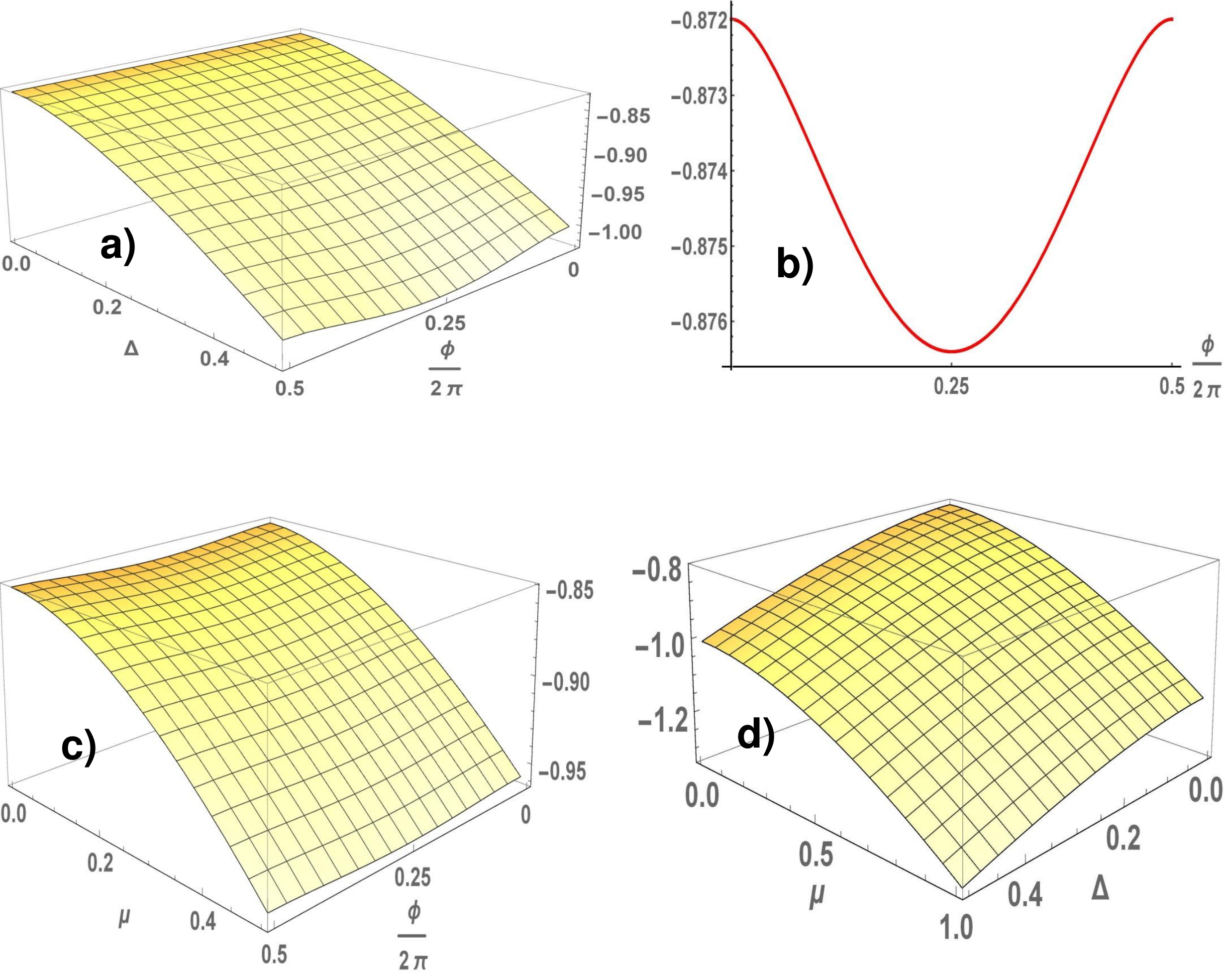}
    \caption{(Color online) A square lattice. The ground state energy per atom as a function of the superconducting order parameter ($\Delta$, $\phi$) at $\mu=0.2$ a), profile at $\Delta =0.2$ b) and as a function of $\mu$ and $\phi$ at $\Delta=0.2$ c). The ground state energy per atom of TSC as function of $\Delta$ and $\mu$ d).
       }
    \label{fig:e1}
\end{figure}

We consider a superconducting state with a spontaneously broken time reversal symmetry  and show that the state of  TSC is realized in the case of symmetry breaking. The energy per cell $E(\phi)$ is determined by the energies
of excitations of fermions $E(\phi)=\sum_{j}\sum_{\mathbf{k},\epsilon_j(\mathbf{k})<\epsilon_F} \epsilon_j(\mathbf{k})$, where $j=1,2$ enumerates the branches of the one-particle excitations $\epsilon_j(\mathbf{k})$
 \begin{equation}
 \epsilon_j(\textbf{k})=(-1)^{j}[(\mu+\cos k_x + \cos k_y)^2 +
     \Delta^2 (\sin^2 k_x + 2 \cos \phi \sin k_x \sin k_y +\sin^2 k_y)]^{1/2},
      \label{eq-2}
     \end{equation}
     where  $\epsilon_F$ is the Fermi energy, $\textbf{ k}=\{k_x,k_y\}$.
The one-particle spectrum Eq(\ref{eq-2}) is symmetric with respect to zero energy, it includes two branches. At half filling a low energy band is filled, therefore the energy of the system $E(\phi)=\sum_{\mathbf{k}} \epsilon_1(\mathbf{k})$ is determined by a low energy branch.  The energy is a periodic function of $\phi$ having periodicity $\pi$ (see in Figs.\ref{fig:e1}a),b),c)). In practice, values of $\Delta,|\mu| <<1$. Therefore, we consider the low energy excitations for $\Delta,|\mu| <1$. It was found numerically that nontrivial solutions $\phi=\pm \pi/2$ correspond to the minimums of the energy for arbitrary values of $\mu \neq 0$ and $\Delta \neq 0$ . The solutions $\phi =0,\pi$, corresponding to a trivial topological phase, satisfy the maximums of the energy (see in Figs \ref{fig:e1}). A trivial topological phase of superconductor is unstable.

\begin{figure}[tp]
    \centering{\leavevmode}
 \includegraphics[width=.7\linewidth]{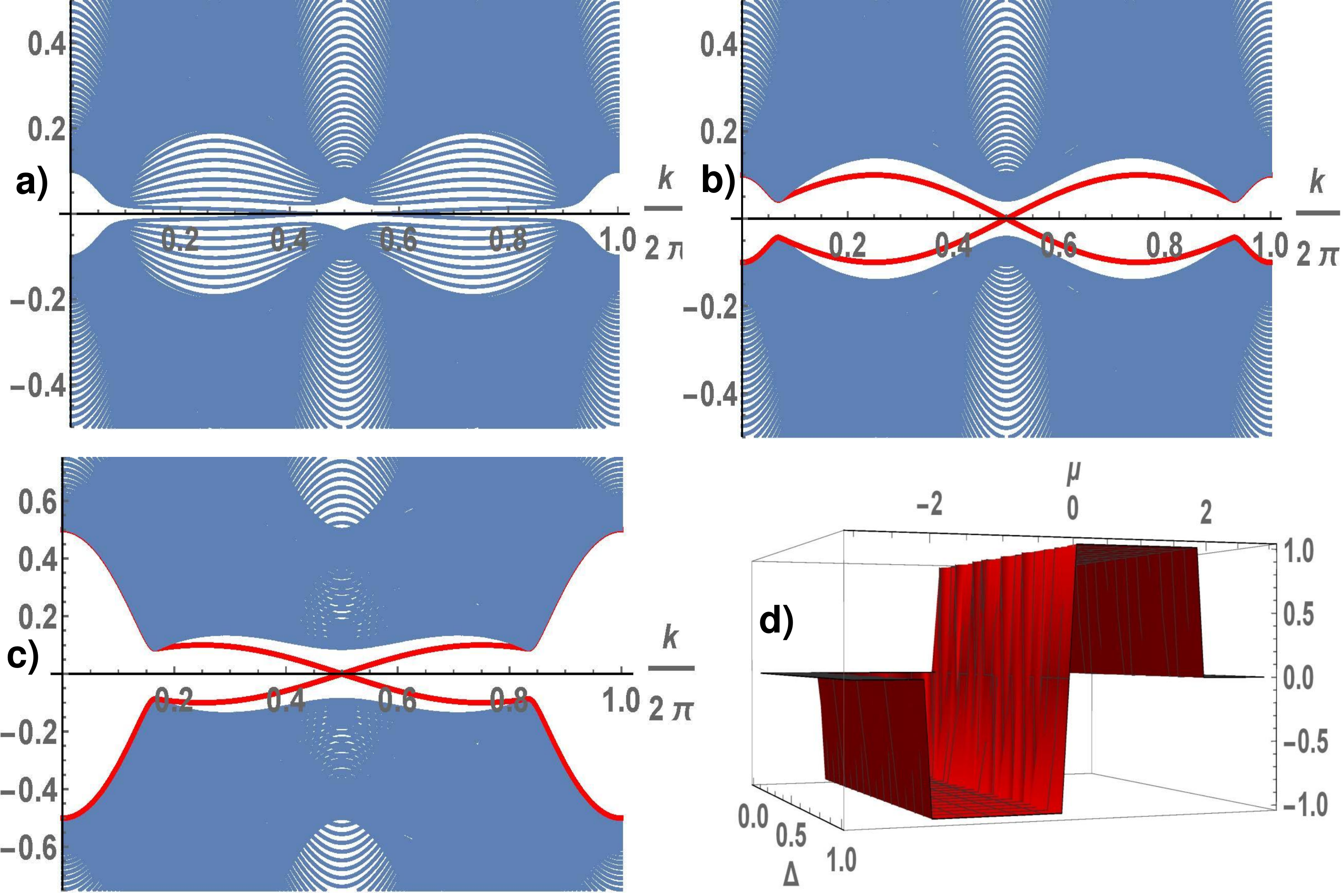}
    \caption{(Color online)
    Low energy spectra of the one-dimensional strip along the \emph{x}-direction as a function of the momentum directed along the edge. The energies are calculated for the following parameter sets: the gapless state with $\phi =0$ at $\mu=0.1$, $\Delta =0.1$ a); the TSC gapped state with $\phi =\frac{\pi}{2}$  at $\mu =0.1$, $\Delta =0.1$ b),  $\mu =0.5$, $\Delta =0.1$ c). In the TSC state the edge modes cross the gap connecting the lower and upper bands and intersect at $k = \pi$. The Chern number as a function of $\Delta$ and $\mu$ d): a strong-pairing trivial phase with $C=0$ at $|\mu|>2$, a weak-pairing topological phase with $C=\pm 1$ at $0<|\mu|<2$.
       }
    \label{fig:sp1}
\end{figure}

The energy band structure of spinless fermions calculated in trivial and nontrivial topological phases is shown in Figs \ref{fig:sp1}.  The gapless excitation spectrum  is realized in the trivial topological phase for arbitrary parameters of the Hamiltonian (see in Figs \ref{fig:sp1} a). In the topological phase the spectrum of the
superconductor is shown for the same parameters  in Fig.\ref{fig:sp1} b). The excitation spectrum varies greatly with $\phi$: in TCS phase the gap in the spectrum of the excitations opens for arbitrary values of $\Delta \neq 0$ and $\mu \neq 0$, the spectrum contains chiral gapless edge modes that populate the bulk gap (see in Figs \ref{fig:sp1} b), c)).
The nonvanishing Chern numbers reveal the nontrivial topological properties of the superconductor. The Chern number is a topological invariant which can be determined for the band isolated from all other bands. In TSC state  the nontrivial Chern number $C= \text{sign}(\mu)$ is realized if the excitation spectrum is gapped at a half-filling:

\begin{equation}
C=\frac{1}{4\pi}\int_{BZ}{\cal B}(\textbf{k})d\textbf{k},\\
{\cal B}(\textbf{k})= -\Delta^2 \frac{\cos k_x +\cos k_y +\mu \cos k_x \cos k_y}
{ \epsilon_2^3(\textbf{k})},
\label{eq-3}
\end{equation}
where integration is carried out over the Brillouin zone.

Let us consider the behavior of the system for arbitrary values of $\mu$ (without limiting $|\mu| <1$), the solutions $\phi= \pm \pi/2$ correspond to a stable state of superconductor for arbitrary values of $\mu$ and $\Delta\neq 0$.
For $\Delta\neq 0$ the gap in the spectrum closes at $\mu =0$ in $\textbf{k}=(0,\pi),(\pi,0)$ and at $\mu =2$ in $\textbf{k}=(\pi,\pi)$, at $\mu =-2$ in $\textbf{k}=(0,0)$.
Following Read and Green \cite{TSC5a}, we expect two phases in the superconductor: a strong-pairing trivial phase with
$C=0$, and a weak-pairing topological phase with $C \neq 0$. According to the calculations of the Chern number (\ref{eq-3}) a strong-pairing trivial phase with $C=0$ is realized in the region $|\mu|>2$  and a weak-pairing topological phase with $C= \text{sign}(\mu)$ is realized at $0<|\mu|<2$  (see in Fig.\ref{fig:sp1} d)).

The nontrivial topological properties of a system are manifested in the existence of chiral gapless edge modes (in two dimensions) or surface  states (in three dimensions) robust against disorder and interactions. The spectrum of TSC includes the chiral edge modes that connect the lower and upper bands. The edge modes determine the charge or spin Hall conductance $\sigma_{xy}$ in the topological state.
Calculations of edge states show that two chiral gapless edge modes are realized in the spectrum of TSC. The amplitude of the wave function of the edge states decreases exponentially away from the boundary, they are localized near the boundaries of the superconductor.

We have determined the phases of the superconducting order parameter in unit cell (see in Figs \ref{fig:Model}). Due to the translational invariance, we have solved the model analytically for the uniform configuration.
As a rule, the energy of different systems has minimum at the same uniform configurations of the gauge fields \cite{Kitaev, Lieb}. In our case topological state, with static configurations $\phi_x(j)$, $\phi_y(j)$ for  nonhomogeneous phases of the superconducting order parameter, is realized at $\phi_x(j)=\phi_x$, $\phi_y(j)=\phi_y$. Numerical calculations of the ground state energy of periodic phases on a square lattice with unit cell containing two and tree atoms confirm these considerations.

\begin{figure}[tp]
    \centering{\leavevmode}
\includegraphics[width=.7\linewidth]{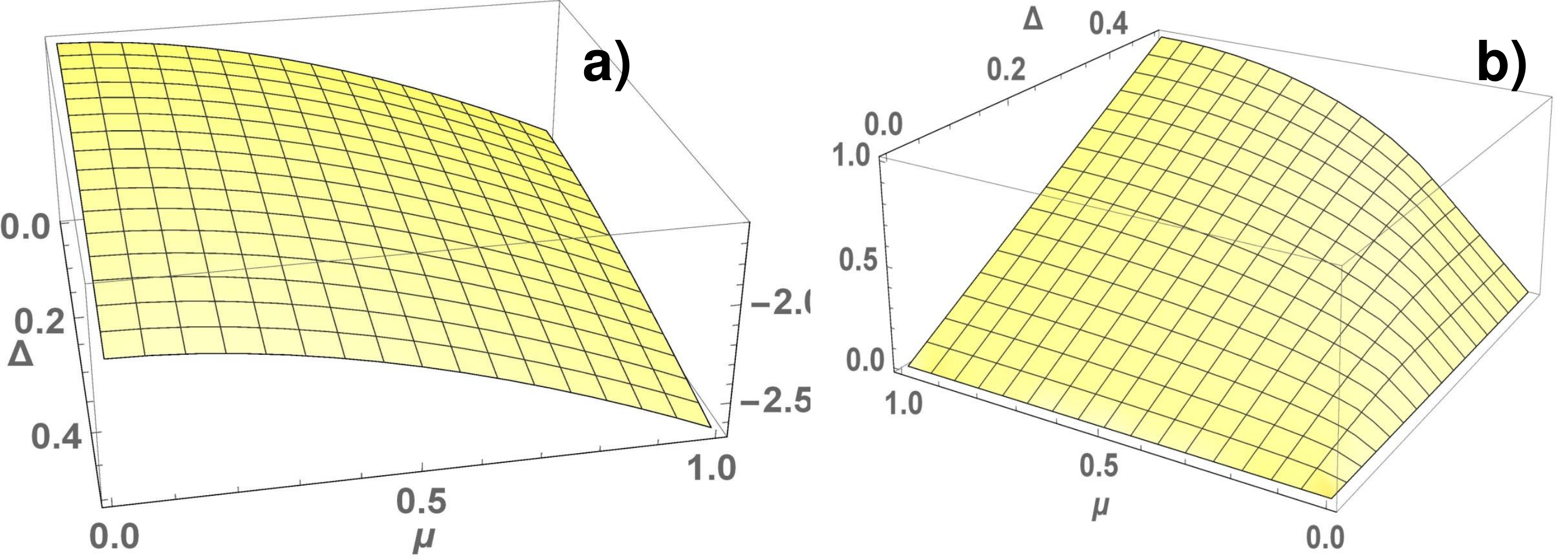}
\caption{(Color online) A centered square lattice. The ground state energy per unit cell a) and the gap in the spectrum b) in the TSC state as functions of $\Delta$ and $\mu$, the energy minimuma  are reached at $\phi=\pm\frac{\pi}{2}$ .
       }
    \label{fig:e2}
\end{figure}

Note that the trivial topological state with $\phi =\varphi=0,\pi$ is unstable in the model  (\ref{eq-H1}) defined on a cubic lattice. The nontrivial solutions for $\phi =0, \varphi= \pm \frac{\pi}{2}$ or  $\phi= \pm \frac{\pi}{2},
\varphi=0$ correspond to the energy minimuma and spontaneously  break the time-reversal symmetry in the 3D system.

\section*{TSC on a centered square lattice }

\begin{figure}[tp]
    \centering{\leavevmode}
 \includegraphics[width=.7\linewidth]{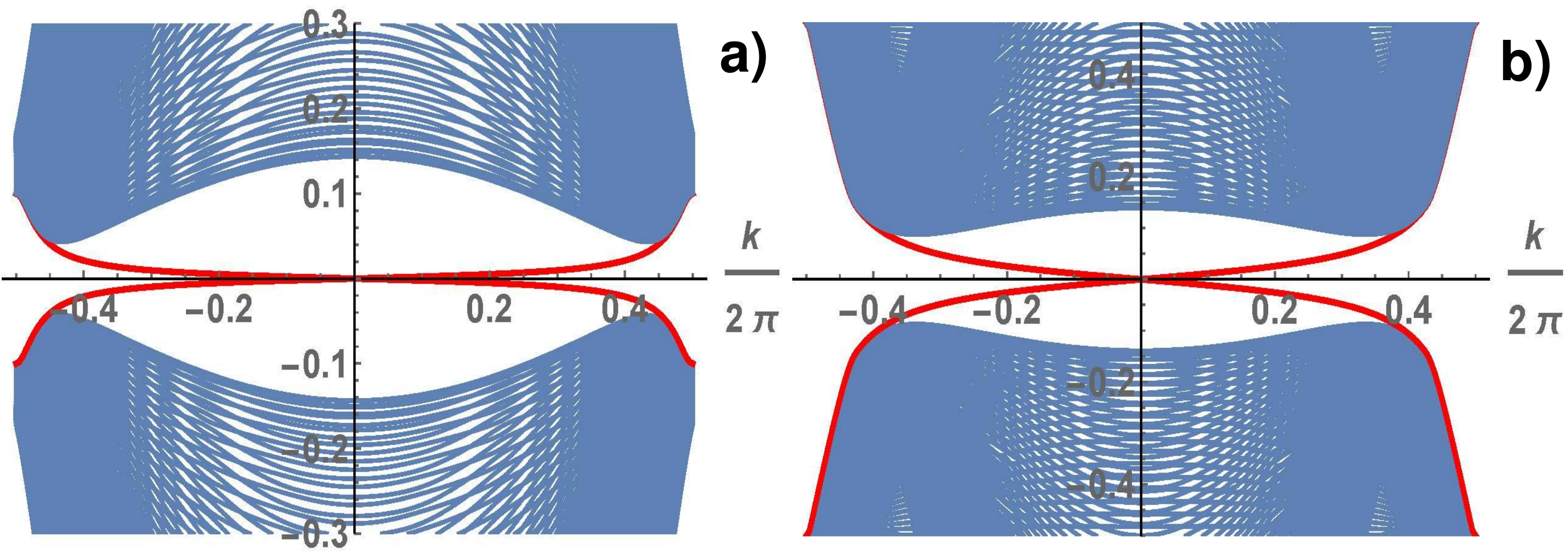}
    \caption{(Color online) Detail of the band structure for a system in the stripe geometry with the same parameters as in Fig.\ref{fig:sp1}; the TSC gapped state with $\phi = \frac{\pi}{2}$  at $\mu =0.1$, $\Delta =0.1$ a), $\mu =0.5$, $\Delta =0.1$ b). Only the low energy excitations are shown, although all four bands were considered in the calculation.
       }
    \label{fig:sp2}
\end{figure}

The spectrum of the one-particle excitations includes four branches $\epsilon_j(\mathbf{k})$ symmetric with respect to zero energy, as result, $E(\phi)=\sum_{j=1,2}\sum_{\mathbf{k}} \epsilon_j(\mathbf{k})$ at half-filling.
According to numerical calculations of the ground state energy for different parameters $\Delta$ and $\mu$ and an arbitrary phase $\phi$, the  energy minima of the superconductor are realized at two nontrivial solutions $\phi= \pm \frac{\pi}{2}$, which correspond to two fold degenerate state of TSC. The behavior of TSC with different crystalline symmetry (centered square and square) is the same. The energy is a $\pi$-periodic function of the phase $\phi$.  The calculation of the ground state energy per cell in the TSC state is shown in Fig.\ref{fig:e2} a). The topological state is determined by the gapped spectrum of fermions for arbitrary $\Delta \neq 0$ and $|\mu|<1$, with the gap illustrated in Fig.\ref{fig:e2} b) (we do not consider an exotic case  $|\mu|>1$). The spectrum of the excitations is gapped for arbitrary parameters of the model, TSC is the stable state of the superconductor. The energy spectra are shown in  Figs \ref{fig:sp2}. The Chern number is equal to one, and the chiral gapless edge modes realize charge or spin Hall conductivity. We can talk about an universal topological behavior of the superconductors.

Taking into account noninteracting spinless fermions, we have considered a new mechanism for the realization of TSCs. The analysis can be generalized to the case of fermions with spin degrees of freedom.
In a small magnetic field $h$ the chemical potentials of the electron subbands with different spins have opposite signs $h\sum_j  n_{j \uparrow}$ and $-h\sum_j  n_{j \downarrow}$. In the case of triplet pairing of the electrons the Hamiltonian of the superconductor is determined by the sum of two decoupled fermion subsystems with their Chern numbers $C_\uparrow =-C_\downarrow$. The spin Chern number $C_{spin}=(C_\uparrow -C_\downarrow)/2$ determines the spin current and the spin Hall conductivity.

\section*{TSC on a hexagonal lattice}

\begin{figure}[tp]
    \centering{\leavevmode}
  \includegraphics[width=\linewidth]{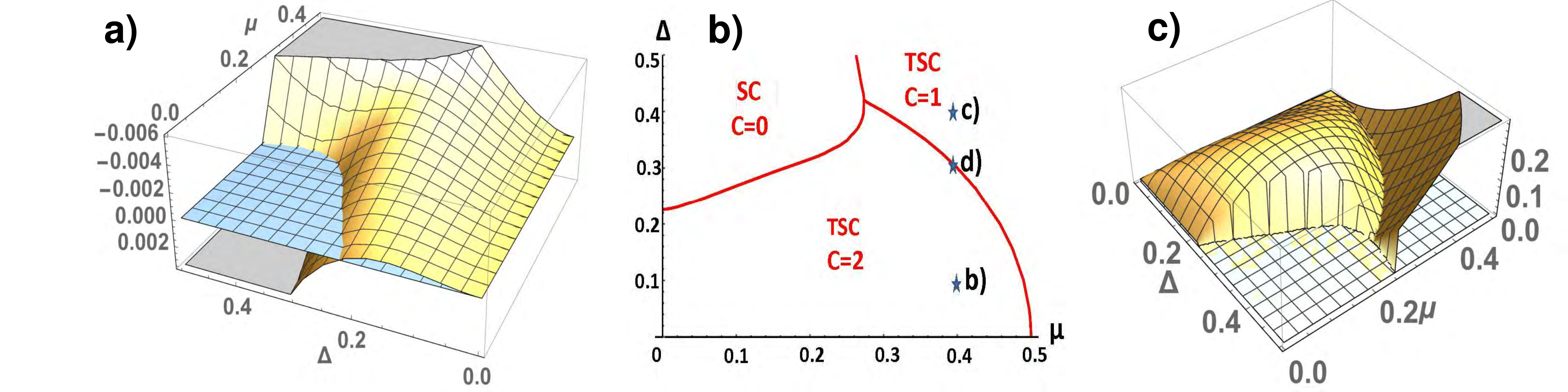}
\caption{(Color online) A hexagonal lattice.  The difference of the energies per unit cell $E(-\frac{\pi}{3},-\frac{2\pi}{3})-E(0,0)$ at $\phi=-\frac{\pi}{3}$, $\varphi=-\frac{2\pi}{3}$ and $\phi=\varphi=0$ as a function of $\mu$ and $\Delta$  a), the ground-state phase diagram in the coordinates  $\Delta$, $\mu$ b):  trivial topological phase with the Chern number equal to 0, and nontrivial topological phases with C=2 and 1. The gap in the spectrum of TSC as a function of $\Delta$ and $\mu$ c). }
\label{fig:e3}
\end{figure}

\begin{figure}[tp]
    \centering{\leavevmode}
\includegraphics[width=.7\linewidth]{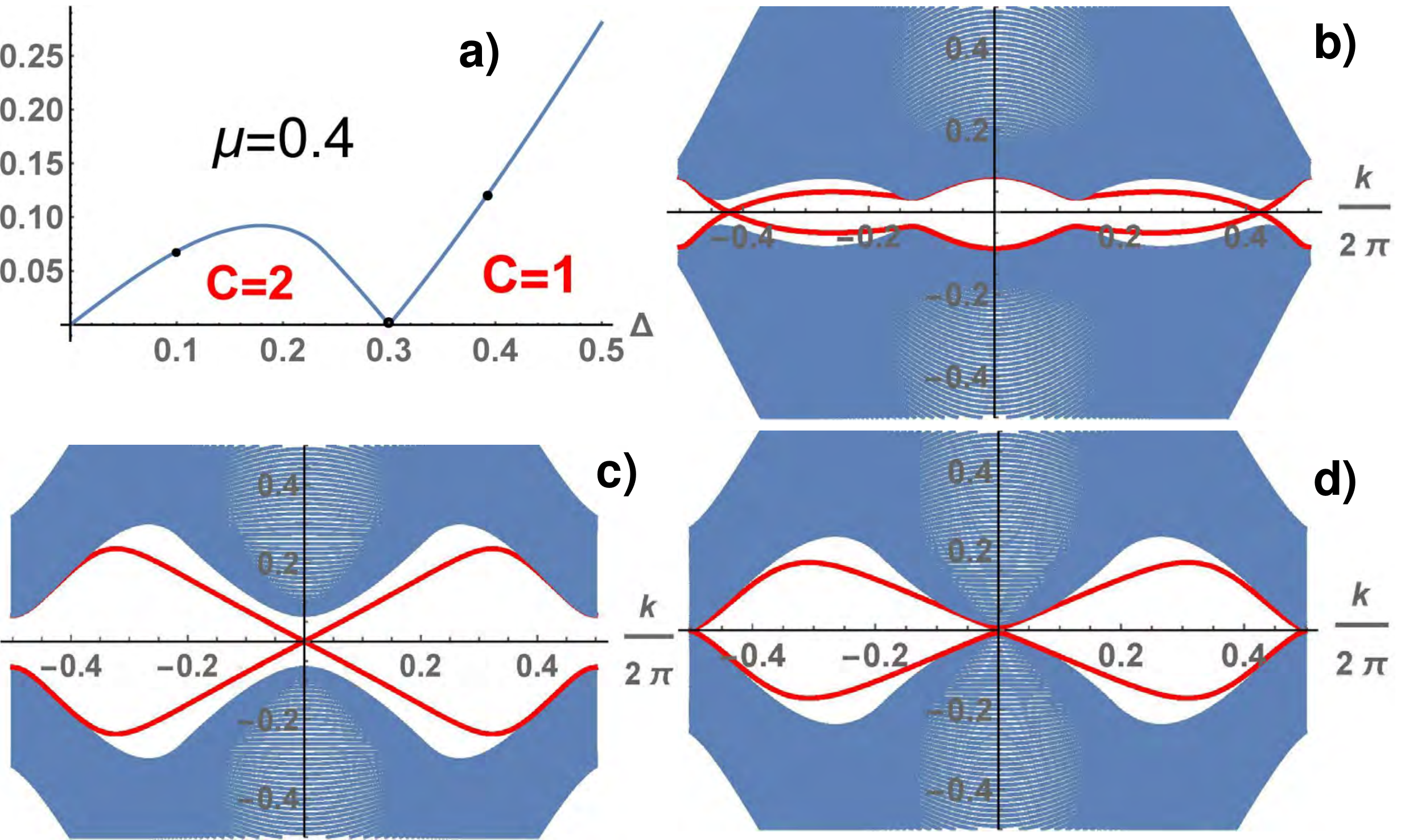}\\d)
\caption{(Color online)
 Evolution of the spectrum of TSC calculated for system with zig-zag boundary along the profile of the gap at $\mu=0.4$ in the points marked in Fig.\ref{fig:e3}b) a) for $\Delta=0.1$ (the state with  C=2) b), $\Delta=0.3$  (the point of the topological phase transition) d), $\Delta=0.4$  (the state with  C=1) c).
 The number of pairs of edge modes is equal to the total Chern number of the bands below the gap, the wave vector directed along the zig-zag boundary.
        }
\label{fig:sp3}
\end{figure}

We can continuously deform the hexagonal lattice by changing the angle between the translation vectors by $\pi/2$, using these perpendicular vectors as a basis of the translation group. The energy per unit cell $E(\phi,\varphi)$ depends on two phases $\phi$ and $\varphi$, it is a periodical function with the period $2\pi$.
The local minima of  $E(\phi,\varphi)$ are reached at $\phi=\varphi=0$ and $\phi=-\frac{\pi}{3}$, $\varphi=-\frac{2 \pi}{3}$ or $\phi=-\frac{2\pi}{3}$ and $\varphi=-\frac{\pi}{3}$. The first solution corresponds to the trivial topological phase, the second one corresponds to the TSC state. Comparing $E(\phi,\varphi)$ calculated for these   phase values (see in Fig.\ref{fig:e3}a)) we determine the ground-state phase diagram of the superconductor in the coordinates of the coupling parameters (see in Fig.\ref{fig:e3}b)). The phase of TSC is stable in the region of small values of $\Delta,\mu$, in practice for arbitrary values. In the TSC phase the fermion spectrum is gapped in contrast to gapless one in the trivial topological phase. The numerical calculations of the gap are shown in  Fig.\ref{fig:e3}c). The gap closes along a line of the topological phase transition in the region of the TSC phase. The closing the bulk gap at certain values of $\Delta,\mu$  leads to the corresponding change of the Chern numbers.
We illustrate the transformation of the spectrum for $\mu=0.4$ and different values of $\Delta=0.1,0.3,0.4$ in Figs \ref{fig:sp3}, where $\Delta=0.3$ is the point of a topological phase transition between phases with the Chern numbers 2 and 1 (at the transition between phases with different Chern numbers, the bulk gap must close). A typical form of the behavior of the spectrum with the zig-zag boundary is shown in the TSC state for different values of $\Delta$ in Figs~\ref{fig:sp3} b), c),d) .

\section*{Conclusions}
In this work we have derived a new approach for description of TSCs.
We have shown that due to nontrivial stable solutions for the phases of the superconducting order parameter, the time-reversal symmetry is spontaneously broken. These solutions for the phases, which determine new topological order parameter in TSC, reveal the nontrivial topological properties of the superconductor. In this context, we have also shown that the TSC state is stable for arbitrary coupling parameters. We have calculated the Chern numbers, studied the chiral gapless edge modes and a Hall conductivity in the TSC state. Considering the square, centered square, hexagonal lattices, we have demonstrated that the nature of the TSC state is independent of the crystalline symmetry of the 2D system.  In contradistinction to tradition approach when the non-trivial topological state of the superconductors arises from the spin-orbit coupling, we have considered the forming of the TSC state as a result of the phase transition with spontaneous breaking of  symmetry. The spontaneous breaking of  time-reversal symmetry (due to a new topological order parameter) is also realized in 3D superconductors.

\section*{Methods}
\subsection*{Canonical functional of topological superconductor}

We show, that the Hamiltonian (\ref{eq-H1}) with different phases of the superconducting order describes the BCS state. The Hamiltonian for spinless fermions is defined as ${\cal H} = {\cal H}_{0}+{\cal H}_{int}$, where ${\cal H}_{int}=-\frac{g}{2} \sum_{<ij>}n_{i}n_{j}$. The interaction between fermions located at the nearest-neighbor lattice sites of a square lattice is additive in the $x-$ and $y-$ directions and is determined by the coupling parameter $g>0$. The Stratonovich transformation maps interacting fermion systems to non-interacting fermions moving in an effective field.
Replacing the sites of fermions interacting on a square lattice in the $x-$direction $i=\alpha-\frac{1}{2}$ and $j=\alpha+\frac{1}{2}$ and in the $y-$direction $i=\gamma-\frac{1}{2}$ and $j=\gamma+\frac{1}{2}$ we define the interaction term as
$\frac{g}{2} (a^\dagger_{\alpha -\frac{1}{2}}a^\dagger_{\alpha +\frac{1}{2}}a_{\alpha +\frac{1}{2}}a_{\alpha -\frac{1}{2}} +
a^\dagger_{\gamma -\frac{1}{2}}a^\dagger_{\gamma+\frac{1}{2}}a_{\gamma +\frac{1}{2}}a_{\gamma-\frac{1}{2}} ) \rightarrow
a^\dagger_{\alpha -\frac{1}{2}}a^\dagger_{\alpha +\frac{1}{2}}\Delta_{x,\alpha}+
\Delta_{x,\alpha}^\ast a_{\alpha+ \frac{1}{2}} a_{\alpha-\frac{1}{2}}+\textbf{}
a^\dagger_{\gamma -\frac{1}{2}}a^\dagger_{\gamma +\frac{1}{2}}\Delta_{y,\beta}+
\Delta_{y,\gamma}^\ast a_{\gamma+ \frac{1}{2}} a_{\gamma-\frac{1}{2}}+
\frac{2}{g}(\Delta_{x,\alpha}^\ast\Delta_{x,\alpha}+\Delta_{y,\gamma}^\ast\Delta_{y,\gamma})$, where \textbf{} $\Delta_{x,\alpha}=\Delta_{\alpha}\exp(i\phi_x)$ and $\Delta_{y,\gamma}=\Delta_{\gamma}\exp(i\phi_y)$.

The canonical functional is defined as
${\cal Z}=\int {\cal D}[\Delta] \int {\cal D}[a^\dagger,a] e^{-S}$, where the action
$S=\frac{2\Delta\Delta^*}{g}+ \int_0^\beta d\tau a^\dagger \partial_\tau a + {\cal H}_{eff}$ with
$ {\cal H}_{eff}=  {\cal H}_0 + [\exp(i\phi_x)\sum_{x-links} \Delta_{\frac{i+j}{2}}a^\dagger_{i}a^\dagger_{j}+ \exp(i\phi_y)\sum_{y-links}\Delta_{\frac{i+j}{2}} a^\dagger_{i}a^\dagger_{j}+h.c.]$.

The BSC Hamiltonian for the interacting spinless fermions is defined as  ${\cal H}_{BCS} = {\cal H}_{0}-\frac{g}{2V} (C_x^+ C_x+C_y^+ C_y)$, where $C_{x,y}=i\sum_{\textbf{k}}\sin k_{x,y} a_\textbf{k} a_{\textbf{-k}}$. Using the Stratanovich transformation for ${\cal H}_{BCS}$ we obtain  the following expression for the effective action
$ S_{eff} =\int_0^\beta d\tau [\sum_\textbf{k} a^+_{\textbf{k}}(\partial_\tau -(\mu+\cos k_x + \cos k_y)) a_{\textbf{k}} +\Delta_x^* C_x+C^+_x\Delta_x+\Delta_y^* C_y+C^+_y\Delta_y + \frac{2}{g}(\Delta_x^*\Delta_x+\Delta_y^*\Delta_y)]$.
We expect that $\Delta_{x,y}$ are independent of $\tau$ because of translational invariance.
We can set $\Delta_x=|\Delta| \exp(i\phi_x)$, $\Delta_y=|\Delta| \exp(i\phi_y)$ and $\phi=\phi_x-\phi_y$,
it is easy to see, that the superconducting order parameter has the following form $\Delta (\textbf{k})=g(\textbf{k},\phi)|\Delta|$, $g(\textbf{k},\phi)=i [\sin k_x + \exp(i\phi)\sin k_y]$.

We can integrate out the fermionic contribution for calculation of the action $\frac{S_{eff}}{\beta}=T\sum_{\textbf{k}}\sum_n \ln [\omega_n^2+\varepsilon_1^2(\textbf{k})]+\frac{4|\Delta|^2}{g}$, where  $\omega_n =T(2n+1)\pi$ are Matsubara frequencies.
In the saddle point approximation the canonical functional ${\cal Z}$ will be dominated by the minimal action $S_{eff}$ that satisfies the following conditions $\partial S_{eff}/\partial \Delta =0$ and $\partial S_{eff}/\partial \phi =0$. The first condition leads to the equation for $|\Delta|$ and $T_c$ or the BCS gap equation
$\frac{4}{g}=T\sum_{\textbf{k}}\sum_{n}\frac{|g(\textbf{k},\phi)|^2}{\omega^2_n+ \varepsilon_1^2(\textbf{k})}$. A numerical calculation of the integral at $T = 0$ showed that its value depends weakly on $ \mu $ for $ 0.01 <\mu <0.2 $, the value of $|\Delta|$ exponentially depends on $-\frac{\nu}{g}$, where $\nu \sim 1$.
The second condition $\sum_{\textbf{k}}\sum_{n}\frac{\partial|g(\textbf{k},\phi)|^2}{\partial \phi}\frac{1}{\omega^2_n+\varepsilon_1^2(\textbf{k})}=0$ determines the topology of the superconducting state. This condition is realized both for trivial $\phi=0,\pi$ and nontrivial $\phi=\pm \frac{\pi}{2}$ solutions. From the numerical calculations of the energy of the superconductor it follows that the trivial solutions correspond to energy maximums whereas the nontrivial solutions are stable and correspond to TSC state (see in Figs \ref{fig:e1}). The topological trivial state of the superconductor is gapless, in the topological state the gap opens and increases with
increasing $\phi$, reaching a maximum value for $\phi = \pi / 2 $. As a result, the energy of the superconductor decreases and reaches a minimum value at $ \phi = \pi /2 $.

\section*{Author contributions statement}

I.K. is an author of the manuscript

\section*{Additional information}

The author declares no competing financial interests.

\end{document}